\documentclass[aip,jmp,reprint,onecolumn,amsmath,amssymb]{revtex4-1}

\usepackage{graphicx}
\usepackage{bm}

\newcommand{\bra}{\langle}
\newcommand{\ket}{\rangle}

\begin{document}

\title{
  Exact description of coalescing eigenstates\\
  in open quantum systems \\
  in terms of microscopic Hamiltonian dynamics
}

\author{Kazuki Kanki}
\author{Savannah Garmon}
\author{Satoshi Tanaka}
\affiliation{Department of Physical Science, 
Osaka Prefecture University, Sakai, Osaka 599-8531, Japan}
\author{Tomio Petrosky}
\affiliation{Center for Complex Quantum Systems, University of Texas at Austin, Texas 78712, USA} 
\affiliation{Institute of Industrial Science, The University of Tokyo, Tokyo 153-8505, Japan}

\date{10 August 2017}

\begin{abstract}

At the exceptional point where two eigenstates coalesce in open quantum systems,
the usual diagonalization scheme breaks down and the Hamiltonian can only be reduced to Jordan block form.
Most of the studies on the exceptional point appearing in the literature introduce a phenomenological effective Hamiltonian that essentially reduces the problem to that of a finite non-Hermitian matrix for which it is straightforward to obtain the Jordan form.
In this paper,
we demonstrate how the microscopic total Hamiltonian of an open quantum system reduces to Jordan block form at an exceptional point in an exact manner that treats the continuum without any approximation
by extending the problem to include eigenstates with complex eigenvalues that reside outside the Hilbert space.
Our method relies on the Brillouin-Wigner-Feshbach projection method according to which we can obtain a finite dimensional effective Hamiltonian that shares the discrete sector of the spectrum with the total Hamiltonian.
Because of the eigenvalue dependence of the effective Hamiltonian due to the dynamical nature of the coupling between the discrete states via the continuum states,
a coalescence of eigenvalues results in the coalescence of the corresponding eigenvectors of the total Hamiltonian,
which means that the system is at an exceptional point.
We also introduce an extended Jordan form basis away from the exceptional point,
which provides an alternative way to obtain the Jordan block at an exceptional point.
The extended Jordan block connects continuously to the Jordan block exactly at the exceptional point implying that the observable quantities are continuous at the exceptional point.

\end{abstract}


\maketitle

\section{Introduction}

While in the traditional formulation of quantum mechanics the generator of time evolution is required to be Hermitian,
in recent years many researchers have found that non-Hermitian operators may be useful in a variety of physical circumstances.\cite{Kadanoff1968,Hatano1996,Hatano1997,Hatano1998,Chalker1997,Narevicius2003,Moiseyev2011}
Besides the fact that eigenvalues of a non-Hermitian operator can be complex,
another remarkable property of non-Hermitian operators compared with Hermitian operators is that eigenvalues and eigenstates show bifurcation and accompanying non-analyticity with regard to parameter dependence.\cite{Serra2001,Klaiman2008a,Klaiman2008b,Cartarius2007,Cartarius2009,Lefebvre2009,Gilary2013,Dembowski2004,Dietz2011,Uzdin2010}
A point in the parameter space where this kind of bifurcation takes place is called an exceptional point.\cite{Kato1976}
At an exceptional point a non-Hermitian degeneracy occurs,
meaning that not only the eigenvalues but also the eigenvectors coalesce, 
and as a result the usual spectral decomposition of the operator fails to hold.\cite{Bhamathi1996,Bohm1997,Antoniou1998,Hernandez2003,Berry2004,Heiss2004,Heiss2012}
The occurrence of non-Hermitian degeneracy is generic for non-Hermitian operators in the sense of the following theorem \cite{Kato1976,Moiseyev1980}: 
\textit{If $S_0$ and $S_1$ are Hermitian and do not commute, then there exists at least one complex value of $\lambda$ for which $S=S_0+\lambda S_1$ is not diagonalizable.}
It is well-known that when a finite dimensional matrix is non-diagonalizable it can only be reduced to Jordan normal form.\cite{Gantmacher1959,Nakanishi1972,Horn2012}

In open quantum systems, the system properties are often described in terms of a phenomenologically obtained finite-dimensional non-Hermitian matrix.\cite{Gunther2007,Demange2012,Rotter2009,Rotter2015}
At the exceptional point,
this matrix is non-diagonalizable and thus takes Jordan block form.
However, the description with this kind of phenomenological Hamiltonian is only approximate,
and the purpose of this paper is to give an exact description of the exceptional point in open quantum systems including continuous spectra.
The coupling between discrete states via the continuum states may lead to coalescence of discrete eigenstates,
while direct coupling always leads to level repulsion.

Generally, the microscopic Hamiltonian for open quantum systems is apparently Hermitian,
meaning that it is Hermitian in the Hilbert space,
but implicitly non-Hermitian in the sense that it may have complex eigenvalues outside of the Hilbert space.\cite{Nakanishi1958,Sudarshan1978,Petrosky1991,Bohm1993}
Hence we can express the Hamiltonian in an extended Hilbert space in which spectral decomposition and resolution of unity can be written in terms of dual eigenvectors that satisfy biorthonormality and bicompleteness, 
except at exceptional points;
in this manner the non-Hermitian nature of the system becomes explicit,
because the total Hamiltonian is represented by a non-Hermitian matrix in a subspace associated with the discrete sector of the spectrum.
The purpose of the present paper is to show that the total Hamiltonian for an open quantum system at an exceptional point can be represented by a Jordan block in a subspace outside of the Hilbert space by introducing pseudo-eigenvectors in addition to eigenvectors as basis vectors.
For simplicity we restrict ourselves to the most common case of exceptional points at which two eigenstates coalesce (EP2).\cite{Heiss2008,Demange2012,Hashimoto2015}
Our formalism can be generalized to higher order exceptional points in a natural way.

We develop our theory on the basis of the Brillouin-Wigner-Feshbach projection method.\cite{Feshbach1958,Feshbach1962,Cohen-Tannoudji1992,Petrosky1997,Rotter2009,Rotter2015,Hatano2013,Hatano2014}
We divide the system into two subsystems:
one of which consists of a finite number of discrete states and the other of which is made of continuum states.
In this formalism the eigenvalue problem of the total Hamiltonian for the discrete eigenvalues is reduced to the eigenvalue problem of an effective Hamiltonian in the finite-dimensional subspace of discrete states.
The distinguishing feature of this effective non-Hermitian Hamiltonian compared with the phenomenological effective Hamiltonian mentioned above is that it depends on the eigenvalue.
In this sense the eigenvalue problem of the effective Hamiltonian is said to be non-linear.
This feature reflects the dynamical nature of the coupling between the discrete states via the continuum states.

The effective Hamiltonian shares the discrete sector of the spectrum with the total Hamiltonian,
and the eigenvector of the total Hamiltonian is uniquely determined from an eigenvector of the effective Hamiltonian.
As a consequence,
coalescence of the eigenstates of the total Hamiltonian at an exceptional point can be understood in terms of the effective Hamiltonian as follows:
When two eigenvalues coalesce,
the corresponding eigenvectors of the effective Hamiltonian also coalesce simply because the corresponding effective Hamiltonian matrices each of which is assigned one of the coalescing eigenvalues become the same.
From a coalescence of eigenvectors of the effective Hamiltonian follows a coalescence of eigenvectors of the total Hamiltonian.
What happens at an exceptional point in terms of the effective Hamiltonian is that each eigenvector of the effective Hamiltonian matrix that is assigned one of the bifurcated eigenvalues coalesces with an eigenvector of the other effective Hamiltonian matrix that is assigned the other of the eigenvalues;
it is not that two eigenvectors of a single effective Hamiltonian matrix coalesce.
In this sense,
an exceptional point of the total Hamiltonian is \emph{not} an exceptional point of the effective Hamiltonian.
Therefore,
the effective Hamiltonian matrix does not have a Jordan block at an exceptional point of the total Hamiltonian,
in contrast to the case of the phenomenological effective Hamiltonian without dependence on the eigenvalue.

The present authors discussed in a separate paper\cite{Tanaka2016} that it is essential to take account of the eigenvalue dependence of the effective Hamiltonian in order to reveal the true nature of the time-symmetry breaking phase transition in open quantum systems.
A time-symmetry breaking phase transition occurs at an exceptional point at which a pair of real eigenvalues coalesce before forming a complex conjugate pair.
This kind of transition takes place also with a phenomenological effective Hamiltonian,
for example,
a parity-time ($\mathcal{PT}$) symmetric Hamiltonian.\cite{Bender2007,Mostafazadeh2010,Bender2002,Garmon2015}
Eigenvalues of a $\mathcal{PT}$-symmetric non-Hermitian Hamiltonian are real in the $\mathcal{PT}$-symmetric phase,
as a result of balanced energy gain and loss.
A $\mathcal{PT}$-symmetric Hamiltonian may show $\mathcal{PT}$-symmetry breaking at an exceptional point where a pair of real eigenvalues turns into a complex conjugate pair.
However, in this case some parts of the system already break time-reversal symmetry even in the $\mathcal{PT}$-symmetric phase.
In this sense the $\mathcal{PT}$-symmetry breaking is an effective breaking of the balance between gain and loss,
and should be distinguished from a genuine time-reversal symmetry breaking as a result of time ordering of dynamical processes.\cite{Petrosky1991}
In any case,
$\mathcal{PT}$-symmetry in a phenomenological effective Hamiltonian is an emergent property of open quantum systems with infinite degrees of freedom resulting from some approximation.
In particular, gain and loss can be balanced at best only on average in a quantum system.\cite{Brody2016}

Recently, Hashimoto \textit{et al.}\cite{Hashimoto2015} pointed out that even away from exceptional points an operator can be represented in terms of an extended Jordan block basis that approaches continuously to the Jordan block at an exceptional point.
This extension of the Jordan block was carried out using finite dimensional matrices with no eigenvalue dependence.
In the present paper,
we generalize the Jordan block further by extending it to describe the coalescence of the eigenvectors of the total Hamiltonian of open quantum systems without introducing any approximation.
This extension serves as an alternative way to obtain the Jordan basis at the exceptional point,
and it reveals the way in which the pseudo-eigenvector emerges out of the coalescing eigenvectors.
Moreover,
the continuous representation of the Hamiltonian emphasizes that the observable quantities are continuous at the exceptional point.

This paper is organized as follows:
In Sec.~\ref{sec:ceigen},
we formulate the complex eigenvalue problem of the total Hamiltonian for open quantum systems in terms of the Brillouin-Wigner-Feshbach formalism.
In Sec.~\ref{sec:EP},
the notion of the exceptional point is introduced.
In Sec.~\ref{sec:pseigen},
we show how to obtain the dual pseudo-eigenvectors,
which together with the dual eigenvectors comprise the Jordan basis at an exceptional point.
In Sec.~\ref{sec:gjordan},
we introduce the extended Jordan basis away from an exceptional point,
and describe how the extended Jordan block approaches the Jordan block exactly at the exceptional point.
Finally, we give concluding remarks in Sec.~\ref{sec:conclusion}.
In Appendix~\ref{app:examples},
we illustrate with simple examples how to construct the Jordan basis at an exceptional point in the Brillouin-Wigner-Feshbach projection method.
In Appendix~\ref{app:Puiseux},
we introduce the Puiseux series,\cite{Kato1976,Moiseyev1980,Hinch1991,Garmon2012,Brody2013} 
which is a fractional power expansion of the eigenvalues and the eigenvectors around an exceptional point.
This expansion is used in Sec.~\ref{sec:gjordan} to describe the limiting behavior of the extended Jordan basis as the system approaches the exceptional point.
Appendix \ref{app:anothersolution} supplement the treatment given in Sec.~\ref{sec:pseigen}.
In Appendix~\ref{app:GJordanFeshbach}, as a complement to Sec.~\ref{sec:gjordan},
we describe how to obtain the extended Jordan basis in terms of the Brillouin-Wigner-Feshbach formalism.

\section{Complex eigenvalue problem of the Hamiltonian on the basis of the Brillouin-Wigner-Feshbach formalism}
\label{sec:ceigen}

In this section, we present a review of the complex eigenvalue problem\cite{Nakanishi1972,Sudarshan1978,Petrosky1991,Bohm1993} of the Hamiltonian for open quantum systems on the basis of the Brillouin-Wigner-Feshbach projection method.\cite{Feshbach1958,Feshbach1962,Cohen-Tannoudji1992,Rotter2009,Rotter2015,Hatano2013,Hatano2014}
We study the quantum dynamics of a single particle with the Hamiltonian of the form

\begin{align}
  H&=H_0+V,
  \\
  H_0&=H_{0\mathrm{d}}+H_{0\mathrm{c}}, 
\end{align}
where $H_{0\mathrm{d}}$ denotes the Hamiltonian in a finite $N$-dimensional subspace,
which is embedded in an infinite dimensional space 
with continuous spectra represented by the Hamiltonian $H_{0\mathrm{c}}$,
and $V$ is the coupling between the states in the two subspaces.
We assume that the Hamiltonian is Hermitian in the Hilbert space.
The finite dimensional subspace consists of spatially localized discrete states such as states in a quantum dot or an impurity.
A discrete state with energy in the continuous spectrum typically decays and results in a resonance state of the total Hamiltonian with a complex eigenvalue.

We consider the complex eigenvalue problem of the Hamiltonian $H$ in the extended Hilbert space,
\begin{subequations}\label{eigeneq}
  \begin{align}
    H|\phi_j\ket=z_j|\phi_j\ket,
    \label{keteigeneq}
    \\
    \bra\tilde\phi_j|H=z_j\bra\tilde\phi_j|,
    \label{braeigeneq}
  \end{align}
\end{subequations}
where $|\phi_j\ket$ and $\bra\tilde\phi_j|$ are 
the right- and left-eigenstates (the eigenket and eigenbra vectors) of $H$ with a common complex eigenvalue $z_j$.
In the complex eigenvalue problem, the left-eigenstate is distinct from the Hermitian conjugate of the right-eigenstate with the same eigenvalue.\cite{Sudarshan1978,Petrosky1991}
Therefore we denote the left-eigenvector with a tilde in order to distinguish it from the Hermitian conjugate of a right-eigenstate.
We can impose the biorthonormality condition on the eigenstates of the Hamiltonian,
i.e.
\begin{equation}
  \bra\tilde\phi_j|\phi_l\ket=\delta_{j,l},
\end{equation}
for eigenstates with discrete eigenvalues,
except at exceptional points [see \eqref{selforthogonal}].
Note that the inner product of an eigenvector with a complex eigenvalue and its Hermitian conjugate vanishes.\cite{Petrosky1991}

We now apply the Brillouin-Wigner-Feshbach projection method to the complex eigenvalue equations.
Let us introduce the projection operator $P$, 
which projects a state into the $N$-dimensional subspace associated with $H_{0\mathrm{d}}$,
and its complement $Q\equiv 1-P$.
They satisfy the following relations:
\begin{equation}
  P^2=P,\quad Q^2=Q,\quad PQ=QP=0,  
\end{equation}
\begin{equation}
  PH_0=H_0P,
\end{equation}
and thus,
\begin{equation}
  PH_0Q=QH_0P=0.
\end{equation}
We assume also that $PVP=0$;
otherwise we redefine $H_0$ to include $PVP$.
Applying both projection operators to \eqref{keteigeneq},
we obtain a set of equations for the $P$- and $Q$-components of an eigenket:
\begin{subequations}
  \begin{align}
    PH_0P|\phi_j\ket+PVQ|\phi_j\ket&=z_jP|\phi_j\ket,
      \label{eigeneqp}
    \\
    QHQ|\phi_j\ket+QVP|\phi_j\ket&=z_jQ|\phi_j\ket.
      \label{eigeneqq}
\end{align}
\end{subequations}
From the second equation \eqref{eigeneqq} we have
\begin{equation}\label{eigenketq}
  Q|\phi_j\ket=\frac{1}{z_j-QHQ}QVP|\phi_j\ket.
\end{equation}
Substituting \eqref{eigenketq} into \eqref{eigeneqp}, we obtain
\begin{equation}\label{eigenHeffket}
  H_\mathrm{eff}(z_j)P|\phi_j\ket=z_jP|\phi_j\ket,
\end{equation}
where $H_\mathrm{eff}(z)$ is the effective Hamiltonian defined by
\begin{equation}\label{Heff}
  H_\mathrm{eff}(z)\equiv PH_0P+\Sigma(z),
\end{equation}
with the self-energy operator $\Sigma(z)$ given by
\begin{equation}\label{selfenergy}
  \Sigma(z)\equiv PVQ\frac{1}{z-QHQ}QVP.
\end{equation}
The operators in the $P$-subspace, $PH_0P, \Sigma(z)$ and $H_\mathrm{eff}(z)$, are represented by 
$N\times N$ matrices.
The operator $\Sigma(z)$ is a function of a complex variable $z$ analytically continued into the second Riemann sheet of the complex energy plane through the cuts on the real axis due to the continuous spectra of $H_{0\mathrm{c}}$.
As a result of the resonance with the continuous spectra,
the effective Hamiltonian becomes non-Hermitian and the eigenvalues can be complex.
Explicit expressions for the effective Hamiltonian and the self-energy are demonstrated in Appendix~\ref{app:examples}.

The eigenvalue problem of the total Hamiltonian in the discrete sector of the spectrum has now been reduced to that of the effective Hamiltonian $H_\mathrm{eff}(z)$:
\begin{equation}\label{det}
  \det\left[ H_\mathrm{eff}(z)-z I_N \right]=0,
\end{equation}
where the effective Hamiltonian $H_\mathrm{eff}(z)$ is represented by a $N\times N$ matrix,
and $I_N$ is the identity matrix of dimension $N$.
Note that in general the eigenvalue equation of the effective Hamiltonian has more than $N$ solutions because of the eigenvalue dependence of the effective Hamiltonian $H_\mathrm{eff}(z)$.\cite{Hatano2014,Tanaka2016,Garmon2017}

The eigenvalue $z_j$ of Eq. \eqref{det} corresponds to the eigenket $P|\phi_j\ket$ of the effective Hamiltonian,
which satisfies Eq.~\eqref{eigenHeffket}.
From the $P$-component we obtain the corresponding $Q$-component by applying \eqref{eigenketq},
and thus we have the eigenket of the total Hamiltonian as $|\phi_j\ket=P|\phi_j\ket+Q|\phi_j\ket$.

Similarly, the $P$-component of the eigenbra $\bra\tilde\phi_j|P$ with an eigenvalue $z_j$ is a solution of the equation,
\begin{equation}\label{eigenHeffbra}
  \bra\tilde\phi_j|PH_\mathrm{eff}(z_j)=z_j\bra\tilde\phi_j|P.
\end{equation}
From the $P$-component we obtain the corresponding $Q$-component as,
\begin{equation}\label{eigenbraq}
  \bra\tilde\phi_j|Q=\bra\tilde\phi_j|PVQ\frac{1}{z_j-QHQ},
\end{equation}
and the eigenbra of the total Hamiltonian is given by $\bra\tilde\phi_j|=\bra\tilde\phi_j|P+\bra\tilde\phi_j|Q$.

In summary,
we can reduce the complex eigenvalue problem of the Hamiltonian in the discrete sector of the spectrum to that of the effective Hamiltonian.
The eigenvalue equation of the effective Hamiltonian is non-linear in the sense that the effective Hamiltonian depends on the eigenvalue.

\section{Exceptional point}
\label{sec:EP}

An exceptional point is a branch point of eigenvalues in the parameter space where not only the eigenvalues but also the eigenstates coalesce.
We denote the set of parameters by ${\bm\kappa}$,
and write the parameter dependence of the Hamiltonian and the effective Hamiltonian as $H({\bm\kappa})$ and $H_\mathrm{eff}({\bm\kappa}; z)$,
respectively.
Let $z_0\equiv z({\bm\kappa}_0)$ denote the eigenvalue at an exceptional point ${\bm\kappa}={\bm\kappa_0}$ that bifurcates into $z_\pm({\bm\kappa})$ for ${\bm\kappa}\neq{\bm\kappa}_0$.
The bifurcated eigenvalues and eigenvectors can be written as \eqref{puiseuxket}--\eqref{puiseuxeval} in terms of square root power expansions with respect to a real parameter $\epsilon$ that measures the deviation from the exceptional point along a curve ${\bm\kappa}(\epsilon)$ in the parameter space\cite{Seyranian2003};
${\bm\kappa}(0)={\bm\kappa}_0$ and the Hamiltonian is given by \eqref{perturbation} in the vicinity of the exceptional point $\epsilon=0$.
For details of this fractional power series,
called the Puiseux series,\cite{Kato1976,Seyranian2003,Hinch1991,Garmon2012}
see Appendix~\ref{app:Puiseux}.
Thus $z_0({\bm\kappa}_0)$ is a double solution of Eq.~\eqref{det} at the exceptional point,
and the non-analytic dependence of eigenvalues on the parameter $\epsilon$ distinguishes the exceptional point from a diabolic point,
where eigenvalues are degenerate in the usual sense and the corresponding eigenvectors remain linearly independent of each other.

In general, two eigenstates with different eigenvalues are orthogonal to each other, 
\begin{equation}\label{orthogonal}
  \bra\tilde\phi_\pm({\bm\kappa})|\phi_\mp({\bm\kappa})\ket=0,
\end{equation}
where $|\phi_\pm({\bm\kappa})\ket$ ($\bra\tilde\phi_\pm({\bm\kappa})|$) are eigenkets (eigenbras) of the Hamiltonian $H({\bm\kappa})$ with corresponding eigenvalues $z_\pm({\bm\kappa})$ for ${\bm\kappa}\neq{\bm\kappa}_0$.
The inner product is continuous with respect to parameters, 
hence by taking the continuous limit of Eq.~\eqref{orthogonal} to the exceptional point ($|\phi_\pm({\bm\kappa})\ket\to|\phi_0({\bm\kappa}_0)\ket$ and  $\bra\tilde\phi_\pm({\bm\kappa})|\to\bra\tilde\phi_0({\bm\kappa}_0)|$ as ${\bm\kappa}\to{\bm\kappa}_0$),
it follows that
\begin{equation}\label{selforthogonal}
  \bra\tilde\phi_0({\bm\kappa}_0)|\phi_0({\bm\kappa}_0)\ket=0,
\end{equation}
where $|\phi_0({\bm\kappa}_0)\ket$ ($\bra\tilde\phi_0({\bm\kappa}_0)|$) is the coalesced eigenket (eigenbra) of the Hamiltonian $H({\bm\kappa}_0)$ at the exceptional point ${\bm\kappa}={\bm\kappa}_0$.
This property is called self-orthogonality.\cite{Narevicius2003,Moiseyev2011}
In other words, the eigenvectors cannot be normalized at an exceptional point in the usual way.
Hence, the self-orthogonality \eqref{selforthogonal} is a necessary condition for a state to bifurcate into two eigenstates under a perturbation.
Conversely,
if the norm of a state does not vanish,
the state cannot bifurcate,
as shown in Appendix~\ref{app:Puiseux}.
Note also that self-orthogonality \eqref{selforthogonal} can never occur for an eigenstate of a Hermitian Hamiltonian within the Hilbert space, 
because the Hermitian conjugate of a right eigenvector $|\phi_j\ket$ in the Hilbert space is the same as the left eigenvector,
i.e. $\bra\tilde\phi_j|=\bra\phi_j|$,
and by definition a vector in the Hilbert space has a non-vanishing finite norm.

The relation \eqref{selforthogonal} can be written as
\begin{align}\label{normpq}
  \bra\tilde\phi_0|P|\phi_0\ket+\bra\tilde\phi_0|Q|\phi_0\ket
  &=\bra\tilde\phi_0|\left[ P+PVQ\frac{1}{(z_0-QHQ)^2}QVP \right]|\phi_0\ket
    \notag
  \\
  &=\bra\tilde\phi_0|P(P-\Sigma'(z_0))P|\phi_0\ket=0,
\end{align}
at ${\bm\kappa}={\bm\kappa}_0$ and the parameter dependence is henceforth omitted.
In the first equality of \eqref{normpq} use has been made of the relations \eqref{eigenketq} and \eqref{eigenbraq},
and $\Sigma'(z)$ is the derivative of the self-energy operator with respect to $z$,
\begin{equation}\label{dSdz}
  \Sigma'(z)=\frac{d\Sigma(z)}{dz}=-PVQ\frac{1}{(z-QHQ)^2}QVP.
\end{equation}
The relation \eqref{normpq} shows that the norm of the $P$-components and that of the $Q$-components exactly cancel with each other at an exceptional point.
Note that when the $P$-subspace is one-dimensional ($N=1$),
Eq.~\eqref{normpq} reduces to $\Sigma'(z_0)=1$, 
which is identical to Eq.~(6) in Ref.~\onlinecite{Garmon2012}.

On the other hand, 
although two eigenstates of the total Hamiltonian with different eigenvalues are orthogonal to each other,
the corresponding eigenstates of the effective Hamiltonian are not in general orthogonal to each other,
\begin{equation}\label{nonorthogonal}
  \bra\tilde\phi_j|P|\phi_l\ket\neq 0,
\end{equation}
where $z_j\neq z_l$.
We can see this fact from the relation
\begin{equation}
  (z_j-z_l)\bra\tilde\phi_j|P|\phi_l\ket
  =\bra\tilde\phi_j|P\left[ H_\mathrm{eff}(z_j)-H_\mathrm{eff}(z_l) \right]P|\phi_l\ket\neq 0,
\end{equation}
which follows from taking the difference of two equations \eqref{eigenHeffket} and \eqref{eigenHeffbra} for the eigenvectors with $z_l$ and $z_j$, respectively.
Therefore the eigenstates of the effective Hamiltonian cannot form an orthogonal basis in the $P$-subspace,
and diagonalization of the effective Hamiltonian does not make sense in the usual way.\cite{note}
Moreover,
as we will show later,
in order to obtain the pseudo-eigenvectors involved in the Jordan basis at an exceptional point for $N\ge 2$,
the eigenvalue problem of the effective Hamiltonian should be generalized to that given by Eq.~\eqref{exteigeneq},
where the eigenvalue in the effective Hamiltonian matrix is fixed at the coalesced eigenvalue $z_0$.

In the special case that the eigenvalue equation of the effective Hamiltonian can be formulated as a quadratic eigenvalue problem,
it can be linearized exactly by doubling the dimension of the matrix to be diagonalized.\cite{Tolstikhin1998,Hatano2014,Ordonez2016}
The resulting generalized eigenvalue equation has the form $(A-\lambda B)|\Psi\ket=0$,
where $A$ and $B$ are matrices both independent of the eigenvalue $\lambda$.
The eigenvalue of the effective Hamiltonian $z$ is a function of $\lambda$,
e.g. $z=-b(\lambda+\lambda^{-1})$ in the case that the continuum is given by the one-dimensional tight-binding model,\cite{Hatano2014}
where $b$ is the transfer integral.
Moreover,
the matrix $B^{-1}A$,
not the effective Hamiltonian itself,
is represented by a matrix containing a Jordan block at an exceptional point.\cite{Garmon2017}
Further,
note that any polynomial eigenvalue problem can be linearized in the sense that any polynomial is a characteristic polynomial of a matrix,
e.g. the Frobenius companion matrix.\cite{Horn2012}
Nonetheless,
the matrices employed in such linearized eigenvalue problems have no simple relation to the original Hamiltonian.
Furthermore,
in the case that the eigenvalue equation of the effective Hamiltonian is non-polynomial,
its linearization may require much more sophisticated manipulation.

\section{Jordan block at an exceptional point}\label{sec:pseigen}

At an exceptional point two eigenstates collapse into one,
and the coalesced eigenvector is supplemented by an associated vector,
called the pseudo-eigenvector,
to span a two-dimensional subspace,
called the generalized eigenspace.
In the generalized eigenspace the Hamiltonian is represented by a Jordan block with the Jordan basis consisting of the eigenvectors and the pseudo-eigenvectors.\cite{Gantmacher1959,Horn2012,Nakanishi1972,Bhamathi1996,Bohm1997,Antoniou1998,Hernandez2003,Demange2012}
In this section,
we assume that the system is at the exceptional point (${\bm\kappa}={\bm\kappa}_0$ in the notation of the previous section) with a coalesced eigenvalue $z_0$,
and show on the basis of the Brillouin-Wigner-Feshbach formalism how to construct the Jordan basis.
The generalized eigenspace associated with the eigenvalue $z_0$ is a subspace in which $(H-z_0)$ is nilpotent;
in the here treated case of two eigenstate coalescence, $(H-z_0)^2=0$ in the subspace.
The Jordan basis in the generalized eigenspace consists of the eigenvectors $|\phi_0\ket, \bra\tilde\phi_0|$ and the pseudo-eigenvectors $|\phi_0^{(1)}\ket, \bra\tilde\phi_0^{(1)}|$, 
which satisfy the relations,
\begin{subequations}
  \begin{align}
    H|\phi_0\ket&=z_0|\phi_0\ket,
    \label{eigenket}
    \\
    (H-z_0)|\phi_0^{(1)}\ket&=c|\phi_0\ket,
    \label{pseigenket}
  \end{align}
\end{subequations}
\begin{subequations}
  \begin{align}
    \bra\tilde\phi_0|H&=z_0\bra\tilde\phi_0|,
    \label{eigenbra}            
    \\
    \bra\tilde\phi_0^{(1)}|(H-z_0)&=c\bra\tilde\phi_0|,
    \label{pseigenbra}
  \end{align}
\end{subequations}
where $c(\neq 0)$ is a constant complex number that can be chosen at one's convenience.
These relations are called the Jordan chain.
As will be demonstrated in Sec.~\ref{subsec:biorthonorm},
we can impose biorthonormality conditions on the eigenvectors and pseudo-eigenvectors,
in addition to the self-orthogonality of the eigenvectors \eqref{selforthogonal},
in order to make them satisfy
\begin{equation}\label{biorthonorm}
  \begin{pmatrix}
    \bra\tilde\phi_0^{(1)}| \\
    \bra\tilde\phi_0|
  \end{pmatrix}
  \begin{pmatrix}
    |\phi_0\ket & |\phi_0^{(1)}\ket
  \end{pmatrix}
  =
  \begin{pmatrix}
    \bra\tilde\phi_0^{(1)}|\phi_0\ket & \bra\tilde\phi_0^{(1)}|\phi_0^{(1)}\ket \\
    \bra\tilde\phi_0|\phi_0\ket       & \bra\tilde\phi_0|\phi_0^{(1)}\ket
  \end{pmatrix}
  =
  \begin{pmatrix}
    1 & 0 \\
    0 & 1
  \end{pmatrix}.
\end{equation}
This relation implies that the eigenvectors and pseudo-eigenvectors form a biorthonormal basis in the generalized eigenspace.
With this basis the Hamiltonian is represented by a Jordan block as
\begin{equation}\label{jordanblock}
  \begin{pmatrix}
    \bra\tilde\phi_0^{(1)}| \\
    \bra\tilde\phi_0|
  \end{pmatrix}
  H
  \begin{pmatrix}
    |\phi_0\ket & |\phi_0^{(1)}\ket
  \end{pmatrix}
  =
  \begin{pmatrix}
    \bra\tilde\phi_0^{(1)}|H|\phi_0\ket & \bra\tilde\phi_0^{(1)}|H|\phi_0^{(1)}\ket \\
    \bra\tilde\phi_0|H|\phi_0\ket       & \bra\tilde\phi_0|H|\phi_0^{(1)}\ket
  \end{pmatrix}
  =
  \begin{pmatrix}
    z_0 & c \\
    0   & z_0
  \end{pmatrix}.
\end{equation}

The arbitrariness of the off-diagonal matrix element $c$ of the Jordan block comes from the fact that the coalesced eigenvectors are self-orthogonal and cannot be normalized in the usual manner.
However,
the physical quantities cannot depend on the value of $c$,
because $c$ is not included in the original Hamiltonian.
This fact can be proved as follows:
From the Jordan basis for a value $c$,
we obtain the Jordan basis for another value $c_1 (\neq 0)$ by multiplying the eigenvectors by $(c_1/c)^{-1/2}$ and the pseudo-eigenvectors by $(c_1/c)^{1/2}$.
In other words,
in the Jordan basis with $c$,
the eigenvectors are proportional to $c^{-1/2}$ and the pseudo-eigenvectors are proportional to $c^{1/2}$,
see for example Eqs. \eqref{pnorm}--\eqref{psa}.
Hence,
the off-diagonal matrix element $c$ is canceled out by the multiplicative factors of the eigenvectors in the representation of the total Hamiltonian with the Jordan basis:
\begin{equation}\label{P0HP0}
P_0 H P_0 = z_0 |\phi_0\ket\bra\tilde\phi_0^{(1)}| + z_0 |\phi_0^{(1)}\ket\bra\tilde\phi_0| + c |\phi_0\ket\bra\tilde\phi_0|,
\end{equation}
where $P_0$ is the projection operator onto the subspace spanned by the Jordan basis vectors.
See also \eqref{norm+} and \eqref{norm-} for ``normalization'' of coalescing eigenvectors.

Let us now assume we have already solved the eigenvalue equations, \eqref{eigenket} and \eqref{eigenbra} in the manner presented in Sec.~\ref{sec:ceigen},
and obtained the exceptional point for which the energy eigenvalues coalesce at $z_0$,
and also the eigenvectors with the eigenvalue $z_0$.
With this assumption,
in the following we solve Eqs. \eqref{pseigenket} and \eqref{pseigenbra} for the pseudo-eigenvectors.
In Sec.~\ref{subsec:pseigeneq} the Brillouin-Wigner-Feshbach projection method is applied to find the equations for the $P$-components of the pseudo-eigenvectors.
We next show that the equations have a solution in Sec.~\ref{subsec:1d} for $N=1$ and in Sec.~\ref{subsec:multidim} for $N\ge 2$.
Finally, in Sec.~\ref{subsec:biorthonorm} we show that the biorthonormality condition \eqref{biorthonorm} can be satisfied.
In Appendix~\ref{app:examples},
we illustrate with simple examples the process to obtain the pseudo-eigenvectors at an exceptional point.

\subsection{Equations for the $P$-components of the pseudo-eigenvectors}\label{subsec:pseigeneq}

Applying the projection operators to \eqref{pseigenket},
we obtain a set of equations for the $P$- and $Q$-components of the pseudo-eigenket:
\begin{subequations}
  \begin{align}
    P(H_0-z_0)P|\phi_0^{(1)}\ket+PVQ|\phi_0^{(1)}\ket&=cP|\phi_0\ket,
      \label{pseigeneqp}
    \\
    QVP|\phi_0^{(1)}\ket+Q(H-z_0)Q|\phi_0^{(1)}\ket&=cQ|\phi_0\ket.
      \label{pseigeneqq}
  \end{align}
\end{subequations}
From the second equation \eqref{pseigeneqq}, we have
\begin{equation}\label{pseigenketq}
  Q|\phi_0^{(1)}\ket=\frac{1}{z_0-QHQ}\left( QVP|\phi_0^{(1)}\ket-cQ|\phi_0\ket \right).
\end{equation}
Substituting \eqref{pseigenketq} into \eqref{pseigeneqp}, we obtain
\begin{align}\label{pseigenketeqp}
  \left( H_\mathrm{eff}(z_0)-z_0 \right)P|\phi_0^{(1)}\ket
  &=c\left( P+PVQ\frac{1}{z_0-QHQ}Q \right)|\phi_0\ket
    \notag
  \\
  &=c\left( P+PVQ\frac{1}{(z_0-QHQ)^2}QVP \right)P|\phi_0\ket
    \notag
  \\
  &=c(P-\Sigma'(z_0))P|\phi_0\ket,
\end{align}
where use has been made of the relation \eqref{eigenketq} in the second equality.

Similarly, the equation for the $P$-component of the pseudo-eigenbra is given by
\begin{align}\label{pseigenbraeqp}
  \bra\tilde\phi_0^{(1)}|P(H_\mathrm{eff}(z_0)-z_0 P)
  &=c\bra\tilde\phi_0|\left( P+Q\frac{1}{z_0-QHQ}QVP \right)
    \notag
  \\
  &=c\bra\tilde\phi_0|P\left( P+PVQ\frac{1}{(z_0-QHQ)^2}QVP \right)
    \notag
  \\
  &=c\bra\tilde\phi_0|P(P-\Sigma'(z_0)),
\end{align}
where use has been made of the relation \eqref{eigenbraq} in the second equality.
The $Q$-component of the pseudo-eigenbra is given in terms of the $P$-component of the pseudo-eigenbra and the $Q$-component of the eigenbra as,
\begin{equation}\label{pseigenbraq}
  \bra\tilde\phi_0^{(1)}|Q=(\bra\tilde\phi_0^{(1)}|PVQ-c\bra\tilde\phi_0|Q)\frac{1}{z_0-QHQ}.
\end{equation}

\subsection{Case of one-dimensional $P$-subspace ($N=1$)}\label{subsec:1d}

In this case,
there is only a single state $|a\ket$ in the $P$-subspace with $\bra a|a\ket=1$ and the projection operator is given by
\begin{equation}\label{P1D}
P=|a\ket\bra a|.
\end{equation}
Then, an operator in the $P$-subspace reduces to a scalar:
e.g. we can identify $H_\mathrm{eff}(z)$ with the single element $\bra a|H_\mathrm{eff}(z)|a\ket$.
Hence, the eigenvalue equation \eqref{det} reduces to a scalar equation,
\begin{equation}\label{dspeq}
  H_\mathrm{eff}(z)-z=0.
\end{equation}
At the exceptional point Eq.~\eqref{dspeq} has a double solution $z=z_0$, 
and \eqref{normpq} implies that
\begin{equation}
  \Sigma'(z_0)=1.
\end{equation}
Therefore \eqref{pseigenketeqp} and \eqref{pseigenbraeqp} are trivially satisfied,
because the multiplication factors on both sides of each equation vanish.

\subsection{Case of multi-dimensional $P$-subspace ($N\ge 2$)}\label{subsec:multidim}

In the equations \eqref{pseigenketeqp} and \eqref{pseigenbraeqp} for the $P$-components of the pseudo-eigenvectors associated with the coalesced eigenvalue $z_0$,
the effective Hamiltonian $H_\mathrm{eff}(z_0)\equiv H_\mathrm{eff}({\bm\kappa}_0; z_0({\bm\kappa}_0))$ is represented by a constant $N\times N$ matrix with its eigenvalue dependence fixed at $z_0$.
Therefore,
now we consider the \emph{linear} eigenvalue problem of the matrix $H_\mathrm{eff}(z_0)$:
\begin{subequations}\label{exteigeneq}
\begin{align}
  H_\mathrm{eff}(z_0)|\psi_j\ket&=z_j|\psi_j\ket,
  \\
  \label{exteigenbraeq}
  \bra\tilde\psi_j|H_\mathrm{eff}(z_0)&=z_j\bra\tilde\psi_j|,
\end{align}
\end{subequations}
where $|\psi_j\ket$ and $\bra\tilde\psi_j|$ are the right- and  left-eigenvectors with an eigenvalue $z_j$.
The eigenvalues $z_j$ are solutions of the equation with respect to $z$ given by
\begin{equation}\label{detext}
  \det\left[ H_\mathrm{eff}(z_0)-z I_N \right]=0,
\end{equation}
which is different from the \emph{non-linear} eigenvalue equation \eqref{det} giving the physical discrete eigenvalues of the total Hamiltonian.
While one of the solutions of Eq.~\eqref{detext} is the eigenvalue $z_0$ of the total Hamiltonian with the eigenvectors given by the $P$-components of those of the total Hamiltonian,
$|\psi_0\ket\propto P|\phi_0\ket$ and $\bra\tilde\psi_0|\propto\bra\tilde\phi_0|P$,
other solutions of Eq.~\eqref{detext} are not eigenvalues of the total Hamiltonian.
In other words,
$z_0$ is in general a \emph{single} solution of Eq.~\eqref{detext},
while it is a \emph{double} solution of Eq.~\eqref{det}.
For an example on this point,
see Appendix~\ref{2d}.
Note that $\bra\tilde\psi_0|\psi_0\ket\propto\bra\tilde\phi_0|P|\phi_0\ket\neq 0$,
while $\bra\tilde\phi_0|\phi_0\ket=0$ (see \eqref{selforthogonal}).

Meanwhile,
we are not aware of any model for which $z_0$ is,
by accident,
a multiple solution of Eq.~\eqref{detext}.
It is perhaps possible that no such physical model exists.
Therefore,
although we have not been able to rule out this possibility,
we restrict ourselves to the case of $z_0$ being a single solution of Eq.~\eqref{detext}.

Because of this restriction,
the eigenspace with the eigenvalue $z_0$ of the effective Hamiltonian,
which is the kernel of the operator $H_\mathrm{eff}(z_0)-z_0 P$ in the $P$-subspace,
is one-dimensional.
In the orthogonal complement of this eigenspace, which is an $(N-1)$-dimensional subspace invariant under the linear transformation $H_\mathrm{eff}(z_0)$, 
the operator $H_\mathrm{eff}(z_0)-z_0 P$ has an inverse,
which we denote by $(H_\mathrm{eff}(z_0)-z_0 P)_\perp^{-1}$.

Finally,
let us now obtain the solution of \eqref{pseigenketeqp}.
According to \eqref{normpq}, the inner product of the right hand side of \eqref{pseigenketeqp} with $\bra\tilde\phi_0|P$ vanishes, 
and thus the ket-vector on the right hand side of \eqref{pseigenketeqp} is in the orthogonal complement of the eigenspace with the eigenvalue $z_0$.
Hence Eq. \eqref{pseigenketeqp} for the $P$-component of the pseudo-eigenket has a solution,
\begin{equation}\label{pseigenketp}
  P|\phi_0^{(1)}\ket=
  c(H_\mathrm{eff}(z_0)-z_0 P)_\perp^{-1}(P-\Sigma'(z_0))P|\phi_0\ket
    +\alpha P|\phi_0\ket,
\end{equation}
where the constant $\alpha$ is arbitrary,
because the term including $\alpha$ vanishes when substituted into the left-hand side of Eq.~\eqref{pseigenketeqp}.
In a similar manner, we obtain the $P$-component of the pseudo-eigenbra as
\begin{equation}\label{pseigenbrap}
  \bra\tilde\phi_0^{(1)}|P=
  c\bra\tilde\phi_0|P(P-\Sigma'(z_0))(H_\mathrm{eff}(z_0)-z_0P)_\perp^{-1}
    +\tilde\alpha\bra\tilde\phi_0|P,
\end{equation}
where $\tilde\alpha$ is an arbitrary constant.

An alternative expression for the $P$-component of the pseudo-eigenvectors is given in Appendix~\ref{app:anothersolution}.

\subsection{Biorthonormality of the Jordan basis}\label{subsec:biorthonorm}

In the previous subsections we obtained the $P$-components of the pseudo-eigenvectors.
By making use of \eqref{pseigenketq} and \eqref{pseigenbraq} we can obtain the $Q$-components of the pseudo-eigenvectors as well.
Thus we have a dual pair of eigenvectors and pseudo-eigenvectors that satisfies Eqs. \eqref{eigenket}--\eqref{pseigenbra}.

Now let us modify the eigenvectors and the pseudo-eigenvectors such that they will satisfy the biorthonormality relations \eqref{biorthonorm}.
The self-orthogonality of the eigenvectors, $\bra\tilde\phi_0|\phi_0\ket=0$, is always true at an exceptional point,
but the self-orthogonality of the pseudo-eigenvectors, $\bra\tilde\phi_0^{(1)}|\phi_0^{(1)}\ket=0$, is not automatically satisfied.
From \eqref{pseigenket} and \eqref{pseigenbra} it follows that
\begin{equation}
  c\bra\tilde\phi_0^{(1)}|\phi_0\ket=c\bra\tilde\phi_0|\phi_0^{(1)}\ket=\bra\tilde\phi_0^{(1)}|(H-z_0)|\phi_0^{(1)}\ket,
\end{equation}
but the value of the inner products $\bra\tilde\phi_0^{(1)}|\phi_0\ket=\bra\tilde\phi_0|\phi_0^{(1)}\ket$ are not necessarily unity.

On the other hand, the pseudo-eigenvectors satisfying \eqref{pseigenket} and \eqref{pseigenbra} are indefinite with regard to the components consisting of the eigenvectors, as indicated by the terms multiplied by the arbitrary constants in \eqref{pseigenketp} and \eqref{pseigenbrap}.
In addition,
an overall multiplication factor can be applied to both $|\phi_0\ket$ and $|\phi_0^{(1)}\ket$, 
or to both $\bra\phi_0|$ and $\bra\phi_0^{(1)}|$.
By taking advantage of these freedoms, we can modify the (pseudo-)eigenvectors to satisfy the biorthonormality relations.
For example, if we modify the (pseudo-)eigenbras as
\begin{align}
  _\mathrm{new}\bra\tilde\phi_0|&=\frac{1}{\bra\tilde\phi_0|\phi_0^{(1)}\ket}\bra\tilde\phi_0|,
  \\
  _\mathrm{new}\bra\tilde\phi_0^{(1)}|&=\frac{1}{\bra\tilde\phi_0^{(1)}|\phi_0\ket}
  \left[ \bra\tilde\phi_0^{(1)}|-\frac{\bra\tilde\phi_0^{(1)}|\phi_0^{(1)}\ket}{\bra\tilde\phi_0|\phi_0^{(1)}\ket}\bra\tilde\phi_0| \right],
\end{align}
then the modified (pseudo-)eigenbras $_\mathrm{new}\bra\tilde\phi_0|,\ _\mathrm{new}\bra\tilde\phi_0^{(1)}|$ and the (pseudo-)eigenkets $|\phi_0\ket,|\phi_0^{(1)}\ket$ satisfy all of the relations \eqref{eigenket}--\eqref{biorthonorm}.

\section{Extended Jordan block basis}\label{sec:gjordan}

The Jordan block is usually introduced only when a matrix cannot be diagonalized.
In that case,
the pseudo-eigenvector compensates for the defect in the spectrum caused by the coalescence of eigenvectors at the exceptional point.
However,
the physical quantities behave continuously at the exceptional point,
and it is favorable to represent the Hamiltonian in a manner where the exceptional point is not treated as a singular point.
For this reason,
Hashimoto \textit{et al.} \cite{Hashimoto2015} introduced an extended Jordan block representation for the generator of time evolution away from the exceptional point that connects continuously to the Jordan block at the exceptional point.
Specifically, in Ref. \onlinecite{Hashimoto2015} the authors considered the Liouvillian dynamics of the density operator and encountered a situation where the eigenvalue dependence of the effective Liouvillian can be neglected,
and hence the effective Liouvillian takes Jordan form at an exceptional point.

In this section we introduce an extended Jordan basis consisting of those eigenstates of the total Hamiltonian in the dual extended Hilbert space that coalesce at the exceptional point,
and demonstrate how the extended Jordan block connects continuously to the Jordan block at the exceptional point.
Whereas here we formally treat this extended Jordan basis in the whole space,
in order to actually apply the formalism presented in this section we have to rely on some method to obtain the eigenvectors of the total Hamiltonian,
e.g. the Brillouin-Wigner-Feshbach method.
We show in Appendix \ref{app:GJordanFeshbach} also how to obtain the extended Jordan basis within the Brillouin-Wigner-Feshbach formalism.

Let $z_+$ and $z_-$ be two non-degenerate discrete eigenvalues of the Hamiltonian that coalesce into an eigenvalue $z_0$ at an exceptional point.
We consider a two dimensional subspace spanned by the eigenvectors $|\phi_+\ket, \bra\tilde\phi_+|$ with the eigenvalue $z_+$ and $|\phi_-\ket, \bra\tilde\phi_-|$ with $z_-$.
We introduce extended pseudo-eigenvectors $|\phi_-^{(1)}\ket$ and $\bra\tilde\phi_+^{(1)}|$ in terms of $|\phi_+\ket, \bra\tilde\phi_+|, |\phi_-\ket$, and $\bra\tilde\phi_-|$ as
\begin{align}
  |\phi_-^{(1)}\ket&\equiv\frac{1}{\bra\tilde\phi_-|\phi_-\ket}|\phi_-\ket+\frac{c}{z_+-z_-}|\phi_+\ket,
    \label{gpseigenket1}
  \\
  \bra\tilde\phi_+^{(1)}|&\equiv\frac{1}{\bra\tilde\phi_+|\phi_+\ket}\bra\tilde\phi_+|+\frac{c}{z_--z_+}\bra\tilde\phi_-|,
    \label{gpseigenbra1}
\end{align}
where $c$ is an arbitrary non-zero constant.
The eigenvectors and the extended pseudo-eigenvectors satisfy the following relations:
\begin{subequations}
  \begin{align}
    H|\phi_+\ket&=z_+|\phi_+\ket,
    \label{geigenket}
    \\
    (H-z_-)|\phi_-^{(1)}\ket&=c|\phi_+\ket,
    \label{gpseigenket}
  \end{align}
\end{subequations}
\begin{subequations}
  \begin{align}
    \bra\tilde\phi_-|H&=z_-\bra\tilde\phi_-|,
    \label{geigenbra}            
    \\
    \bra\tilde\phi_+^{(1)}|(H-z_+)&=c\bra\tilde\phi_-|,
    \label{gpseigenbra}
  \end{align}
\end{subequations}
\begin{equation}\label{gbiorthonorm}
  \begin{pmatrix}
    \bra\tilde\phi_+^{(1)}| \\
    \bra\tilde\phi_-|
  \end{pmatrix}
  \begin{pmatrix}
    |\phi_+\ket & |\phi_-^{(1)}\ket
  \end{pmatrix}
  =
  \begin{pmatrix}
    \bra\tilde\phi_+^{(1)}|\phi_+\ket & \bra\tilde\phi_+^{(1)}|\phi_-^{(1)}\ket \\
    \bra\tilde\phi_-|\phi_+\ket    & \bra\tilde\phi_-|\phi_-^{(1)}\ket
  \end{pmatrix}
  =
  \begin{pmatrix}
    1 & 0 \\
    0 & 1
  \end{pmatrix}.
\end{equation}
Thus the Hamiltonian is represented by an extended Jordan block in the two dimensional subspace:
\begin{equation}
  \begin{pmatrix}
    \bra\tilde\phi_+^{(1)}| \\
    \bra\tilde\phi_-|
  \end{pmatrix}
  H
  \begin{pmatrix}
    |\phi_+\ket & |\phi_-^{(1)}\ket
  \end{pmatrix}
  =
  \begin{pmatrix}
    \bra\tilde\phi_+^{(1)}|H|\phi_+\ket & \bra\tilde\phi_+^{(1)}|H|\phi_-^{(1)}\ket \\
    \bra\tilde\phi_-|H|\phi_+\ket    & \bra\tilde\phi_-|H|\phi_-^{(1)}\ket
  \end{pmatrix}
  =
  \begin{pmatrix}
    z_+ & c \\
    0   & z_-
  \end{pmatrix}.
\end{equation}
Note that here the normalization of the eigenvectors, 
i.e. $\bra\tilde\phi_+|\phi_+\ket$ and $\bra\tilde\phi_-|\phi_-\ket$, 
is arbitrary,
because the eigenvectors $\bra\tilde\phi_+|$ and $|\phi_-\ket$ are not used as biorthonormal basis vectors in this representation.
Note also that the physical quantities do not depend on the value of $c$,
which can be proved in a similar manner as described in the paragraph including Eq.~\eqref{P0HP0}.

In order to examine the behavior of the extended pseudo-eigenvectors \eqref{gpseigenket1} and \eqref{gpseigenbra1} when the system approaches the exceptional point,
we rewrite them as
\begin{align}
  |\phi_-^{(1)}\ket&=\frac{c}{z_--z_+}\left[ |\phi_-\ket-|\phi_+\ket \right]
  +\left[ \frac{1}{\bra\tilde\phi_-|\phi_-\ket}-\frac{c}{z_--z_+} \right]|\phi_-\ket,
  \label{pseigenketoff}
  \\
  \bra\tilde\phi_+^{(1)}|&=\frac{c}{z_+-z_-}\left[ \bra\tilde\phi_+|-\bra\tilde\phi_-| \right]
  +\left[ \frac{1}{\bra\tilde\phi_+|\phi_+\ket}-\frac{c}{z_+-z_-} \right]\bra\tilde\phi_+|.
  \label{pseigenbraoff}
\end{align}
If the two eigenvalues became degenerate in the usual sense,
the second term on the right hand side of each of \eqref{pseigenketoff} and \eqref{pseigenbraoff} would diverge.
This is because $z_+-z_-\to 0$ but $\bra\tilde\phi_\pm|\phi_\pm\ket$ does not vanish at the degeneracy point.
Hence, the diabolic point, where the usual degeneracy of the eigenvalue occurs,
is a singular point of the extended Jordan form.
On the other hand, at an exceptional point the norm of the eigenstate vanishes [the self-orthogonality, see \eqref{selforthogonal}],
and as a result the two singularities in each of the second terms of \eqref{pseigenketoff} and \eqref{pseigenbraoff} cancel each other.
Consequently,
the pseudo-eigenvectors of the extended Jordan basis converge to the pseudo-eigenvectors of the Jordan form exactly at the exceptional point,
as we show below.

By making use of the Puiseux series expansions for the eigenvalues and the eigenvectors (see Appendix \ref{app:Puiseux}),
which describe how the eigenstate at an exceptional point bifurcates into two eigenstates,
we can show that the following limits exist:
\begin{align}
  \lim_{\epsilon\to 0}\frac{1}{z_+-z_-}\left[ |\phi_+\ket-|\phi_-\ket \right]&\equiv|\phi_0'\ket,
  \label{dphiket}
  \\
  \lim_{\epsilon\to 0}\frac{1}{z_+-z_-}\left[ \bra\tilde\phi_+|-\bra\tilde\phi_-| \right]
  &\equiv\bra\tilde\phi_0'|
  \label{dphibra}
\end{align}
where $\epsilon$ measures the deviation away from the exceptional point in the parameter space [see \eqref{perturbation}],
and that the eigenvectors can be ``normalized'' so that 
\begin{align}
  \bra\tilde\phi_-|\phi_-\ket&=(z_--z_+)/c+N_2(z_--z_+)^2/c+\mathcal{O}\left( (z_--z_+)^3 \right),
  \label{norm+}
  \\
  \bra\tilde\phi_+|\phi_+\ket&=(z_+-z_-)/c+N_2(z_+-z_-)^2/c+\mathcal{O}\left( (z_+-z_-)^3 \right).
  \label{norm-}
\end{align}
These formulae can be derived by making use of \eqref{puiseuxket}--\eqref{puiseuxeval} and noting that $z_+-z_-\propto\sqrt{\epsilon}$.
The two eigenstates coalesce at the exceptional point as
\begin{align}
  \lim_{\epsilon\to 0}|\phi_+\ket&=\lim_{\epsilon\to 0}|\phi_-\ket\equiv|\phi_0\ket,
  \\
  \lim_{\epsilon\to 0}\bra\tilde\phi_+|&=\lim_{\epsilon\to 0}\bra\tilde\phi_-|\equiv\bra\tilde\phi_0|,
\end{align}
where $|\phi_0\ket$ and $\bra\tilde\phi_0|$ are the eigenvectors at the exceptional point.
Under the limiting behaviors \eqref{dphiket}--\eqref{norm-}, 
the extended pseudo-eigenvectors,
\eqref{pseigenketoff} and \eqref{pseigenbraoff},
converge to the following limits at the exceptional point:
\begin{align}
  |\phi_0^{(1)}\ket=c|\phi_0'\ket-cN_2|\phi_0\ket,
  \label{lpseigenket}
  \\
  \bra\tilde\phi_0^{(1)}|=c\bra\tilde\phi_0'|-cN_2\bra\tilde\phi_0|.
  \label{lpseigenbra}
\end{align}

By taking the limit ($\epsilon\to 0$) to the exceptional point in \eqref{geigenket}--\eqref{gbiorthonorm},
we obtain \eqref{eigenket}--\eqref{biorthonorm}.
Therefore the Jordan basis at the exceptional point has been obtained as a limit of the extended Jordan basis away from the exceptional point.
This fact implies that the observable quantities change smoothly when passing through the exceptional point.

\section{Concluding remarks}\label{sec:conclusion}

The implicit non-Hermiticity of open quantum systems has recently attracted much interest.
However,
non-Hermitian components are often introduced phenomenologically without fully taking account of their origin.
Precisely speaking, the Hamiltonian for an open quantum system is Hermitian in the Hilbert space;
nonetheless,
it may have eigenstates with complex eigenvalues outside of the Hilbert space,
and then it can be represented by a non-Hermitian matrix in the extended Hilbert space.
Moreover,
this representation gives us important physical information such as the decay rate of an unstable state,
determined as the imaginary part of a complex eigenvalue.

In this paper we have shown that the total Hamiltonian for an open quantum system at an exceptional point is represented by a Jordan block in a subspace spanned by basis vectors that reside outside of the Hilbert space.
We described how to obtain a Jordan basis that consists of dual eigenvectors and associated dual pseudo-eigenvectors with use of the Brillouin-Wigner-Feshbach formalism.
In this approach the complex eigenvalue problem of the total Hamiltonian is reduced to that of the effective Hamiltonian,
which depends on the eigenvalue itself.
Because of the eigenvalue dependence,
a coalescence of eigenvalues results in the coalescence of both the corresponding eigenvectors of the effective Hamiltonian and those of the total Hamiltonian,
although the effective Hamiltonian does not have a Jordan block
in contrast to the case of a phenomenological effective Hamiltonian.
Our method
allows us to write the Jordan block of the \emph{total} Hamiltonian without introducing any phenomenological argument or approximation

We then showed that the Jordan block at an exceptional point can be obtained by taking the appropriate limit of an extended Jordan block basis that can be written away from an exceptional point in a subspace spanned by two eigenstates that bifurcate from the coalesced eigenstate at the exceptional point.
This is an extension of the formalism introduced in Ref.\onlinecite{Hashimoto2015} for finite dimensional matrices to the case of operators in an extended Hilbert space,
fully incorporating the continuum.
An insight gained from the study of the extended Jordan basis is that the pseudo-eigenvectors are necessary in order to eliminate the divergence due to self-orthogonality of the coalesced eigenvectors in the spectral decomposition.

While in this paper we dealt with the problem in the Hamiltonian formalism for the wave function, 
we have previously found exceptional points also in the Liouvillian formalism for the density operator,\cite{Hashimoto2015,Hashimoto2016a,Hashimoto2016b,Hashimoto2016c,Kanki2016}
where the effective Liouvillian is conventionally called the collision operator.\cite{Petrosky1997}
For related results, see also Ref. \onlinecite{Kimura2002,Qian2000,Am-Shallem2015,Am-Shallem2016}.
However our studies concerning the exceptional point have so far been restricted to the level of an approximate effective Liouvillian,
in which the eigenvalue dependence is neglected.
In the future,
we aim to investigate the eigenvalue problem of the effective Liouvillian in which the eigenvalue dependence is retained.

\appendix

\section{How to obtain the Jordan basis with examples}
\label{app:examples}

In this Appendix we illustrate how to obtain the Jordan basis vectors at an exceptional point by examining two models investigated in Ref.~\onlinecite{Tanaka2016} as examples.
With this basis the total Hamiltonian is represented by a Jordan block as \eqref{jordanblock}.

\subsection{A single discrete state coupled with a continuum}

In one of the models a single discrete state $|a\ket$ couples with a continuum as,
\begin{subequations}\label{model1}
  \begin{align}
    H_0&=\varepsilon_a |a\ket\bra a|+\int d\bm{k}\,\varepsilon_{\bm{k}}|\bm{k}\ket\bra\bm{k}|,
    \\
    V&=\int d\bm{k}\, v(\bm{k}) \left( |a\ket\bra\bm{k}|+|\bm{k}\ket\bra a| \right),
  \end{align}
\end{subequations}
where $\bm{k}$ is the wave-vector designating a continuum state with energy eigenvalue $\varepsilon_{\bm{k}}$.
With the projection operator $P=|a\ket\bra a|$,
the effective Hamiltonian and the self-energy are scalars given by
\begin{align}
  H_\mathrm{eff}(z)=\varepsilon_a+\Sigma(z),
  \\
  \Sigma(z)=\int d\bm{k} \frac{v^2(\bm{k})}{z-\varepsilon_{\bm{k}}}.
\end{align}

We specify the model following Ref.~\onlinecite{Tanaka2016} such that the continuum is given by a one-dimensional free particle with the energy dispersion $\varepsilon_k=\hbar^2 k^2/2m$,
and $v(k)$ is a constant $\alpha$  in the range $-k_\mathrm{c}<k<k_\mathrm{c}$ with a wave-number cutoff $k_\mathrm{c}$.
We use the units in which $\hbar=1, k_\mathrm{c}=1$ and $2m=1$.
The self-energy is given by $\Sigma(z)=\alpha^2 \sigma(z)$ with
\begin{equation}\label{sigmaz}
  \sigma(z)=1-i\frac{\pi}{2\sqrt{z}}.
\end{equation}
The eigenvalue equation of the effective Hamiltonian \eqref{dspeq} reduces to the cubic equation\cite{5x5}
\begin{equation}\label{cubic}
  z(z-\varepsilon_a - \alpha^2)^2+\frac{\pi^2\alpha^4}{4}=0.
\end{equation}
We found an exceptional point at $\varepsilon_a=\varepsilon_{\mathrm{c},1}\equiv-3(\pi\alpha^2/4)^{2/3}-\alpha^2$,
where a pair of real eigenvalues turns into a complex conjugate pair as $\varepsilon_a$ increases while the other parameters are fixed (see Fig.~2 in Ref.~\onlinecite{Tanaka2016}).
The coalesced eigenvalue at the exceptional point is $z_0=-(\pi\alpha^2/4)^{2/3}$.

From the $P$-components,
proportional to $|a\ket$ or $\bra a|$,
of the Jordan basis vectors follow the complementary $Q$-components according to \eqref{eigenketq}, \eqref{eigenbraq}, \eqref{pseigenketq} and \eqref{pseigenbraq}.
In order for the biorthonormality conditions \eqref{biorthonorm} to be satisfied by the Jordan basis vectors thus obtained,
the $P$-components of them must satisfy the relations,
\begin{equation}\label{pnorm}
  \bra\tilde\phi_0|a\ket\bra a|\phi_0\ket=-\frac{2}{c\,\Sigma''(z_0)},
\end{equation}
and
\begin{equation}
  \bra\tilde\phi_0|a\ket\bra a|\phi_0^{(1)}\ket+\bra\tilde\phi_0^{(1)}|a\ket\bra a|\phi_0\ket
  =\frac{2}{3} \frac{\Sigma'''(z_0)}{(\Sigma''(z_0))^2},
\end{equation}
where a prime on the selfenergy $\Sigma(z)$ means differentiation with respect to $z$,
see \eqref{dSdz}.
For example, a solution of these equations is given by
\begin{equation}
  \bra a|\phi_0\ket=\bra\tilde\phi_0|a\ket=\left( \frac{-2}{c\,\Sigma''(z_0)} \right)^{1/2}
\end{equation}
and
\begin{equation}\label{psa}
  \bra a|\phi_0^{(1)}\ket=\bra\tilde\phi_0^{(1)}|a\ket=\frac{(2c)^{1/2}}{3} \frac{\Sigma'''(z_0)}{(-\Sigma''(z_0))^{3/2}}.
\end{equation}

\subsection{Two discrete states coupled with a continuum}\label{2d}

In the second model two discrete states $|a\ket$ and $|b\ket$ couple with the continuum as,
\begin{subequations}\label{model2}
  \begin{align}
    H_0&=\varepsilon_a |a\ket\bra a| + \varepsilon_b |b\ket\bra b| + \int d\bm{k}\,\varepsilon_{\bm{k}}|\bm{k}\ket\bra\bm{k}|,
    \\
    V&=\int d\bm{k}\, v_a(\bm{k}) \left( |a\ket\bra\bm{k}|+|\bm{k}\ket\bra a| \right)
      +\int d\bm{k}\, v_b(\bm{k}) \left( |b\ket\bra\bm{k}|+|\bm{k}\ket\bra b| \right).
  \end{align}
\end{subequations}
With the projection operator $P=|a\ket\bra a|+|b\ket\bra b|$,
the effective Hamiltonian and the self-energy are represented by $2\times 2$ matrices as,
\begin{align}
  H_\mathrm{eff}(z)&=
  \begin{pmatrix}
    \varepsilon_a+\Sigma_{aa}(z) & \Sigma_{ab}(z) \\
    \Sigma_{ab}(z) & \varepsilon_b+\Sigma_{bb}(z)
  \end{pmatrix},
  \\
  \Sigma_{cd}(z)&=\int d\bm{k} \frac{v_c(\bm{k}) v_d(\bm{k})}{z-\varepsilon_{\bm{k}}},
\end{align}
where $c,d\in\{a,b\}$.

We specify the model in the same manner,
and use the same units,
as in the previous subsection.
Here we set $v_a(k)$ and $v_b(k)$ to be constants $\alpha_a$ and $\alpha_b$, respectively, in the range $-k_\mathrm{c}<k<k_\mathrm{c}$.
The components of the self-energy are given by $\Sigma_{cd}(z)=\alpha_c \alpha_d \sigma(z)$ ($c,d\in\{a,b\}$),
where $\sigma(z)$ is given by \eqref{sigmaz}.
The eigenvalue equation of the effective Hamiltonian \eqref{det} is given by
\begin{equation}\label{evaeq2x2}
  \left( z-\varepsilon_a-\Sigma_{aa}(z) \right) \left( z-\varepsilon_b-\Sigma_{bb}(z) \right)
  -\left( \Sigma_{ab}(z) \right)^2=0.
\end{equation}
The equation \eqref{evaeq2x2} reduces to the quintic equation\cite{5x5}
\begin{equation}\label{quintic}
  4z\{(z-\varepsilon_a-\alpha_a^2)(z-\varepsilon_b-\alpha_b^2)-\alpha_a^2 \alpha_b^2\}^2
  +\pi^2\{(\alpha_a^2 + \alpha_b^2)z-(\alpha_b^2 \varepsilon_a + \alpha_a^2 \varepsilon_b)\}^2 =0.
\end{equation}

\begin{figure}[ht]
\includegraphics[width=8cm,clip]{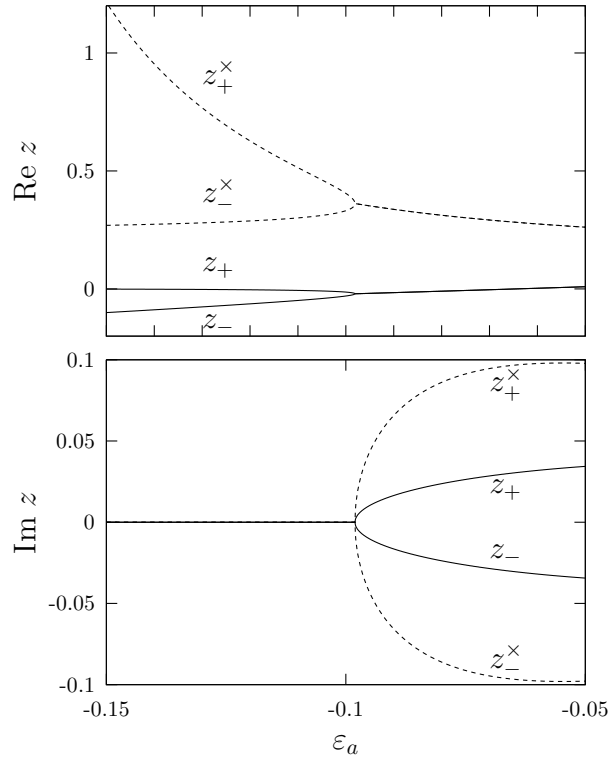}
\caption{\label{exteigen}
The eigenvalues of the $2\times 2$ matrices $H_\mathrm{eff}(z_+(\varepsilon_a))$ and $H_\mathrm{eff}(z_-(\varepsilon_a))$ in the vicinity of the exceptional point.
The solid curves represent genuine eigenvalues $z_+(\varepsilon_a)$ and $z_-(\varepsilon_a)$,
and the dashed curves represent $z_+^\times(\varepsilon_a)$ and $z_-^\times(\varepsilon_a)$,
which are not eigenvalues of the total Hamiltonian.
}
\end{figure}

Now we focus our attention on the exceptional point at $\varepsilon_a=-0.099\equiv\tilde\varepsilon_{\mathrm{c},1}, \varepsilon_b=0.2$ and $\alpha_a=\alpha_b=0.1$,
where a pair of real eigenvalues turns into a complex conjugate pair as $\varepsilon_a$ increases while the other parameters are fixed (see Fig.~4 in Ref.~\onlinecite{Tanaka2016}).
In Fig.~\ref{exteigen} we show the eigenvalues of the $2\times 2$ matrices $H_\mathrm{eff}(z_+(\varepsilon_a))$ and $H_\mathrm{eff}(z_-(\varepsilon_a))$ in the vicinity of the exceptional point,
where the eigenvalue in each effective Hamiltonian matrix is fixed at one of the eigenvalues $z_\pm(\varepsilon_a)$  bifurcated from the eigenvalue $z_0\equiv z_0(\tilde\varepsilon_{\mathrm{c},1})$ at the exceptional point.
One of the eigenvalues of the matrix $H_\mathrm{eff}(z_+(\varepsilon_a))$ ($H_\mathrm{eff}(z_-(\varepsilon_a))$),
$z_+(\varepsilon_a)$ ($z_-(\varepsilon_a)$),
is an eigenvalue of the total Hamiltonian,
while the other $z_+^\times(\varepsilon_a)$ ($z_-^\times(\varepsilon_a)$) is not an eigenvalue of the total Hamiltonian.
The extraneous eigenvalues $z_+^\times(\varepsilon_a)$ and $z_-^\times(\varepsilon_a)$ converge as $\varepsilon_a\to\tilde\varepsilon_{\mathrm{c},1}$ to a common value,
which we denote by $z_0^\times$ ($\neq z_0$).
Each of the eigenvectors of $H_\mathrm{eff}(z_+(\varepsilon_a))$ coalesces with one of the eigenvectors of $H_\mathrm{eff}(z_-(\varepsilon_a))$ at the exceptional point $\varepsilon_a=\tilde\varepsilon_{\mathrm{c},1}$,
just because the two effective Hamiltonian matrices become the same.

At the exceptional point,
the $2\times 2$ matrix representing the effective Hamiltonian $H_\mathrm{eff}(z_0)$ can be diagonalized,
and has two different eigenvalues,
$z_0$ and $z_0^\times$.
Hence,
there are two linearly independent eigenvectors of $H_\mathrm{eff}(z_0)$.
One of them is the $P$-components of the coalesced eigenvector of the total Hamiltonian with the eigenvalue $z_0$:
$P|\phi_0\ket$ or $\bra\phi_0|P$.
The other is related to the $P$-component of the pseudo-eigenvector:
Eqs. \eqref{pseigenketp} and \eqref{pseigenbrap} can be rewritten in this case as
\begin{eqnarray}
  P|\phi_0^{(1)}\ket=\frac{c}{z_0^\times -z_0}|\psi_0^\times\ket\frac{\bra\tilde\psi_0^\times|(P-\Sigma'(z_0))P|\phi_0\ket}{\bra\tilde\psi_0^\times|\psi_0^\times\ket}+\alpha P|\phi_0\ket,
\end{eqnarray}
and
\begin{equation}
  \bra\tilde\phi_0^{(1)}|P=\frac{c}{z_0^\times -z_0}\frac{\bra\tilde\phi_0|P(P-\Sigma'(z_0))|\psi_0^\times\ket}{\bra\tilde\psi_0^\times|\psi_0^\times\ket}\bra\tilde\psi_0^\times|+\tilde\alpha\bra\tilde\phi_0|P,
\end{equation}
respectively,
where $|\psi_0^\times\ket$ and $\bra\tilde\psi_0^\times|$ are the eigenvectors with the eigenvalue $z_0^\times\ (\neq z_0)$ of the effective Hamiltonian $H_\mathrm{eff}(z_0)$ satisfying the eigenvalue equations \eqref{exteigeneq} with $z_j=z_0^\times$.
From the $P$-components of the Jordan basis vectors follow the complementary $Q$-components according to \eqref{eigenketq}, \eqref{eigenbraq}, \eqref{pseigenketq} and \eqref{pseigenbraq}.

\section{Puiseux series}\label{app:Puiseux}

In this Appendix we show that the eigenvalues and the eigenvectors of the Hamiltonian can be expanded in a fractional power series\cite{Kato1976,Seyranian2003,Hinch1991,Garmon2012} about an exceptional point with respect to a real parameter $\epsilon$, where $\epsilon$ measures the deviation away from the exceptional point in the parameter space.
This fractional power series is known as the Puiseux series.
For this purpose, we write the eigenvalue equation of the Hamiltonian near an exceptional point in the form\cite{Hinch1991}
\begin{equation}\label{perturbation}
  (H_\mathrm{EP}+\epsilon H')|\phi\ket=z|\phi\ket,
\end{equation}
where $H_\mathrm{EP}$ is the Hamiltonian at the exceptional point,
and it is assumed that $[H_\mathrm{EP},H']\neq 0$.
Then, the eigenvector and the eigenvalue are expanded in powers of $\epsilon$ away from the exceptional point:
\begin{align}
  |\phi\ket&=|\phi_0\ket+|\bar\phi_1\ket+|\bar\phi_2\ket+\cdots,
  \\
  z&=z_0+\bar{z}_1+\bar{z}_2+\cdots,
\end{align}
where each quantity with a bar contains a successively increasing power of $\epsilon$.
The first terms $|\phi_0\ket$ and $z_0$ are the coalesced eigenvector and eigenvalue at the exceptional point and satisfy
\begin{equation}
  H_\mathrm{EP}|\phi_0\ket=z_0|\phi_0\ket,
\end{equation}
which is the zero-th order equation in \eqref{perturbation}.
Moreover, the eigenvector $|\phi_0\ket$ is self-orthogonal with the left-eigenvector $\bra\tilde\phi_0|$ having the same eigenvalue $z_0$:
$\bra\tilde\phi_0|\phi_0\ket=0$ [see \eqref{selforthogonal}].

The lowest order equation resulting from the inner product of Eq.~\eqref{perturbation} with $\bra\tilde\phi_0|$ is given by
\begin{equation}\label{phi1}
  \epsilon\bra\tilde\phi_0|H'|\phi_0\ket=\bar{z}_1\bra\tilde\phi_0|\bar\phi_1\ket,
\end{equation}
where use has been made of the fact that $\bra\tilde\phi_0|\phi_0\ket=0$.
The left-hand side of \eqref{phi1} is assumed not to vanish,
otherwise the eigenstate at $\epsilon=0$ does not bifurcate under the perturbation.
Then, \eqref{phi1} implies that $\bar{z}_1|\bar\phi_1\ket$ is proportional to $\epsilon$ and thus each of $\bar{z}_1$ and $|\bar\phi_1\ket$ has a fractional power of $\epsilon$.
Then it follows that the first order equation in \eqref{perturbation} is
\begin{equation}\label{first}
  (H_\mathrm{EP}-z_0)|\bar\phi_1\ket=\bar{z}_1|\phi_0\ket.
\end{equation}
Equation \eqref{first} implies that $|\bar\phi_1\ket$ is a pseudo-eigenvector and that $|\bar\phi_1\ket$ and $\bar{z}_1$ have the same power of $\epsilon$.
From this fact and Eq.~\eqref{phi1} it follows that $|\bar\phi_1\ket$ and $\bar{z}_1$ are both proportional to $\epsilon^{1/2}$.

Therefore we conclude that the eigenvalues and the eigenvectors are expanded near an exceptional point in a power series with half-integer exponents.
Because of the two-valuedness of the square root function,
the single eigenstate at the exceptional point bifurcates into two eigenstates as we move away from the exceptional point.

For comparison,
let us consider the case that $|\phi_0\ket$ and $\bra\tilde\phi_0|$ are the eigenvectors with a non-degenerate eigenvalue $z_0$ of an unperturbed Hamiltonian $H_\mathrm{ND}$ representing a system not at an exceptional point.
In this case, $\bra\tilde\phi_0|\phi_0\ket\neq 0$, 
and we have the lowest order equations for (\ref{perturbation}) with $H_\mathrm{EP}$ replaced by $H_\mathrm{ND}$,
\begin{align}
  \epsilon\bra\tilde\phi_0|H'|\phi_0\ket=\bar{z}_1\bra\tilde\phi_0|\phi_0\ket,
  \\
  (H_\mathrm{ND}-z_0)|\bar\phi_1\ket=-\epsilon H'|\phi_0\ket,
\end{align}
instead of \eqref{phi1} and \eqref{first}.
Hence, each of $\bar{z}_1$ and $|\bar\phi_1\ket$ is proportional to $\epsilon$,
and we have the usual integer power expansions with respect to $\epsilon$.

To summarize, the eigenvalues and the eigenvectors are expanded about an exceptional point in the following forms:
\begin{align}
  |\phi_\pm\ket&=|\phi_0\ket\pm\sqrt{\epsilon}|\phi_1\ket+\epsilon|\phi_2\ket+\cdots,
  \label{puiseuxket}
  \\
  \bra\tilde\phi_\pm|&=\bra\tilde\phi_0|\pm\sqrt{\epsilon}\bra\tilde\phi_1|+\epsilon\bra\tilde\phi_2|+\cdots,
  \label{puiseuxbra}
  \\
  z_\pm&=z_0\pm\sqrt{\epsilon}z_1+\epsilon z_2+\cdots,
  \label{puiseuxeval}
\end{align}
where $\sqrt{\epsilon}$ is defined to be $\sqrt{\epsilon}>0$ for $\epsilon>0$ 
and $\sqrt{\epsilon}=i\sqrt{|\epsilon|}$ for $\epsilon<0$.

\section{Alternative expression for the $P$-component of the pseudo-eigenvectors}\label{app:anothersolution}

In this Appendix we give an expression for the $P$-components of the pseudo-eigenvectors at an exceptional point alternative to that given in Sec.~\ref{subsec:multidim},
i.e. \eqref{pseigenketp} and \eqref{pseigenbrap},
in the case of a multi-dimensional $P$-subspace.

Let $z_+$ and $z_-$ denote the eigenvalues that bifurcate from $z_0$.
We take the difference of the two equations for the eigenket \eqref{eigenHeffket} for $z_+$ and $z_-$,
divide it by $(z_+-z_-)$, and take the limit to the exceptional point;
as a result, we obtain
\begin{equation}\label{eigenketpdiff}
  (H_\mathrm{eff}(z_0)-z_0)\lim_{\epsilon\to 0}\frac{P|\phi_+\ket-P|\phi_-\ket}{z_+-z_-}
  =(P-\Sigma'(z_0))P|\phi_0\ket.
\end{equation}
Comparing \eqref{pseigenketeqp} and \eqref{eigenketpdiff}, 
we obtain a solution for the $P$-component of the pseudo-eigenket as
\begin{equation}
  P|\phi_0^{(1)}\ket=c\lim_{\epsilon\to 0}\frac{P|\phi_+\ket-P|\phi_-\ket}{z_+-z_-}
    +\beta P|\phi_0\ket,
\end{equation}
where $\beta$ is an arbitrary constant.
In a similar manner, we obtain the $P$-component of the pseudo-eigenbra as
\begin{equation}
  \bra\tilde\phi_0^{(1)}|P=c\lim_{\epsilon\to 0}\frac{\bra\tilde\phi_+|P-\bra\tilde\phi_-|P}{z_+-z_-}
    +\tilde\beta\bra\tilde\phi_0|P
\end{equation}
where $\tilde\beta$ is an arbitrary constant.

It turns out that the solution in this Appendix can also be obtained by taking the $P$-components of \eqref{lpseigenket} and \eqref{lpseigenbra}.

\section{Extended Jordan block with the Brillouin-Wigner-Feshbach formalism}\label{app:GJordanFeshbach}

In this Appendix we present a method to obtain the extended pseudo-eigenvectors to represent the Hamiltonian in the extended Jordan block form away from the exceptional point.
We assume the eigenvalues and eigenvectors satisfying \eqref{geigenket} and \eqref{geigenbra} have already been obtained.
In the following we solve the equations for the pseudo-eigenvectors, \eqref{gpseigenket} and \eqref{gpseigenbra}:
In Sec.~\ref{subsec:gpseigeneq} the Brillouin-Wigner-Feshbach projection method is applied.
We next show that the equations have a solution in Sec.~\ref{subsec:g1d} for $N=1$ and in Sec.~\ref{subsec:gmultidim} for $N\ge 2$.
Finally, in Sec.~\ref{subsec:gbiorthonorm} we show that the biorthonormality condition \eqref{gbiorthonorm} can be satisfied.

\subsection{Equations for the $P$-components of the pseudo-eigenvectors}\label{subsec:gpseigeneq}

Applying projection operators to \eqref{gpseigenket}, 
we obtain a set of equations for the $P$- and $Q$-components of a pseudo-eigenket:
\begin{subequations}
  \begin{align}
    P(H_0-z_-)P|\phi_-^{(1)}\ket+PVQ|\phi_-^{(1)}\ket=cP|\phi_+\ket,
    \label{gpseigeneqp}
    \\
    QVP|\phi_-^{(1)}\ket+Q(H-z_-)Q|\phi_-^{(1)}\ket=cQ|\phi_+\ket.
    \label{gpseigeneqq}
  \end{align}
\end{subequations}
From the second equation \eqref{gpseigeneqq}, we have
\begin{equation}\label{gpseigenketq}
  Q|\phi_-^{(1)}\ket=\frac{1}{z_--QHQ}(QVP|\phi_-^{(1)}\ket-cQ|\phi_+\ket).
\end{equation}
Substituting \eqref{gpseigenketq} into \eqref{gpseigeneqp}, 
we obtain an equation for the $P$-component of the pseudo-eigenket:
\begin{align}\label{gpseigenketeqp}
  (H_\mathrm{eff}(z_-)-z_-)P|\phi_-^{(1)}\ket
  &=c\left( P+PVQ\frac{1}{z_--QHQ}Q \right)|\phi_+\ket
  \notag
  \\
  &=c\left( P+PVQ\frac{1}{z_--QHQ}\frac{1}{z_+-QHQ}QVP \right)P|\phi_+\ket,
\end{align}
where use has been made of the relation \eqref{eigenketq} for $z_j=z_+$ in the second equality.

Similarly, the equation for the $P$-component of the pseudo-eigenbra is given by
\begin{align}\label{gpseigenbraeqp}
  \bra\tilde\phi_+^{(1)}|P(H_\mathrm{eff}(z_+)-z_+)
  &=c\bra\tilde\phi_-|\left( P+Q\frac{1}{z_+-QHQ}QVP \right)
  \notag
  \\
  &=c\bra\tilde\phi_-|\left( P+PVQ\frac{1}{z_--QHQ}\frac{1}{z_+-QHQ}QVP \right),
\end{align}
where use has been made of the relation \eqref{eigenbraq} for $z_j=z_-$ in the second equality.
The $Q$-component of the pseudo-eigenbra is given 
in terms of the $P$-component of the pseudo-eigenbra and the $Q$-component of the eigenbra as,
\begin{equation}\label{gpseigenbraq}
  \bra\tilde\phi_+^{(1)}|Q=(\bra\tilde\phi_+^{(1)}|PVQ-c\bra\tilde\phi_-|Q)\frac{1}{z_+-QHQ}.
\end{equation}

\subsection{Case of one-dimensional $P$-subspace ($N=1$)}\label{subsec:g1d}

In this case, the projection operator is given by \eqref{P1D},
and operators in the $P$-subspace reduce to scalars.
Hence, $H_\mathrm{eff}(z)-z=0$ for $z=z_\pm$, and the relation
\begin{equation}\label{orthomp}
  \bra\tilde\phi_-|\phi_+\ket=\bra\tilde\phi_-|\left( P+PVQ\frac{1}{z_--QHQ}\frac{1}{z_+-QHQ}QVP \right)|\phi_+\ket=0
\end{equation}
results in
\begin{equation}
  P+PVQ\frac{1}{z_--QHQ}\frac{1}{z_+-QHQ}QVP =0.
\end{equation}
Therefore \eqref{gpseigenketeqp} and \eqref{gpseigenbraeqp} are trivially satisfied,
because the multiplication factors on both sides of each equation vanish.

\subsection{Case of multi-dimensional $P$-subspace ($N\ge 2$)}\label{subsec:gmultidim}

We assume that each of the eigenvalues $z_-$ and $z_+$ is non-degenerate.
Then the eigenspace with the eigenvalue $z_\pm$, which is the kernel of the operator $H_\mathrm{eff}(z_\pm)-z_\pm P$,
is one-dimensional.
In the orthogonal complement of this eigenspace, which is an $(N-1)$-dimensional subspace, 
the operator $H_\mathrm{eff}(z_\pm)-z_\pm P$ has an inverse,
and we denote it by $(H_\mathrm{eff}(z_\pm)-z_\pm P)_\perp^{-1}$.

According to \eqref{orthomp}, the inner product of the right hand side of \eqref{gpseigenketeqp} with $\bra\tilde\phi_-|P$ vanishes, 
and thus the ket-vector on the right hand side of \eqref{gpseigenketeqp} is in the orthogonal complement of the eigenspace.
Hence Eq. \eqref{gpseigenketeqp} for the $P$-component of the pseudo-eigenket has the solution,
\begin{equation}\label{gpseigenketp}
  P|\phi_-^{(1)}\ket=
  c(H_\mathrm{eff}(z_-)-z_-P)_\perp^{-1}\left( P+PVQ\frac{1}{z_--QHQ}Q\right)|\phi_+\ket
    +\alpha_- P|\phi_-\ket,
\end{equation}
where $\alpha_-$ is an arbitrary constant.
In a similar manner, we obtain the $P$-component of the pseudo-eigenbra as
\begin{equation}\label{gpseigenbrap}
  \bra\tilde\phi_+^{(1)}|P=
  c\bra\tilde\phi_-|\left( P+Q\frac{1}{z_+-QHQ}QVP \right)(H_\mathrm{eff}(z_+)-z_+P)_\perp^{-1}
    +\tilde\alpha_+\bra\tilde\phi_+|P,
\end{equation}
where $\tilde\alpha_+$ is an arbitrary constant.

\subsection{Biorthonormality of the extended Jordan basis}\label{subsec:gbiorthonorm}

In the previous subsection we obtained the $P$-components of the pseudo-eigenvectors.
By making use of \eqref{gpseigenketq} and \eqref{gpseigenbraq} we obtain the $Q$-components of the pseudo-eigenvectors.
Thus we have a dual pair of eigenvectors and pseudo-eigenvectors that satisfies Eqs. \eqref{geigenket}--\eqref{gpseigenbra}.

Now let us make the eigenvectors and the pseudo-eigenvectors satisfying the biorthonormality relations \eqref{gbiorthonorm}.
The orthogonality between the eigenvectors with different eigenvalues, $\bra\tilde\phi_-|\phi_+\ket=0$, 
holds automatically,
but the orthogonality between the pseudo-eigenvectors associated with different eigenvalues, 
$\bra\tilde\phi_+^{(1)}|\phi_-^{(1)}\ket=0$, is a condition imposed by hand.
On the other hand, from \eqref{gpseigenket} and \eqref{gpseigenbra} it follows that
\begin{equation}
  (z_+-z_-)\bra\tilde\phi_+^{(1)}|\phi_-^{(1)}\ket=c\left( \bra\tilde\phi_+^{(1)}|\phi_+\ket-\bra\tilde\phi_-|\phi_-^{(1)}\ket \right).
\end{equation}
Hence, $\bra\tilde\phi_+^{(1)}|\phi_-^{(1)}\ket=0$ is satisfied if $\bra\tilde\phi_+^{(1)}|\phi_+\ket=\bra\tilde\phi_-|\phi_-^{(1)}\ket$ holds.

Meanwhile, the pseudo-eigenvectors satisfying \eqref{gpseigenket} and \eqref{gpseigenbra} are indefinite with regard to the eigenvector components, 
indicated by the terms multiplied by the arbitrary constants in \eqref{gpseigenketp} and \eqref{gpseigenbrap}.
We can choose these arbitrary constants so that $\bra\tilde\phi_+^{(1)}|\phi_+\ket=\bra\tilde\phi_-|\phi_-^{(1)}\ket$ holds.
In addition an overall multiplication factor can be applied to both $|\phi_+\ket$ and $|\phi_-^{(1)}\ket$, 
or to both $\bra\phi_-|$ and $\bra\phi_+^{(1)}|$.
By taking advantage of these freedoms, we can modify the (pseudo-)eigenvectors to satisfy $\bra\tilde\phi_+^{(1)}|\phi_+\ket=\bra\tilde\phi_-|\phi_-^{(1)}\ket=1$.


\begin{thebibliography}{99}
%
\bibitem{Kadanoff1968}
L. P. Kadanoff and J. Swift, Phys. Rev. \textbf{165}, 310 (1968).
%
\bibitem{Hatano1996}
N. Hatano and D. R. Nelson, Phys. Rev. Lett. \textbf{77}, 570 (1996).
%
\bibitem{Hatano1997}  
N. Hatano and D. R. Nelson, Phys. Rev. B \textbf{56}, 8651 (1997).
%
\bibitem{Hatano1998}
N. Hatano and D. R. Nelson, Phys. Rev. B \textbf{58}, 8384 (1998).
%
\bibitem{Chalker1997}
J. T. Chalker and Z. J. Wang, Phys. Rev. Lett. \textbf{79}, 1797 (1997).
%
\bibitem{Narevicius2003}
E. Narevicius, P. Serra, and N. Moiseyev, Europhys. Lett. \textbf{62}, 789 (2003).
%
\bibitem{Moiseyev2011}
N. Moiseyev, \textit{Non-Hermitian Quantum Mechanics} (Cambridge University Press, 2011).
%
\bibitem{Serra2001}
P. Serra, S. Kais, and N. Moiseyev, Phys. Rev. A \textbf{64}, 062502 (2001).
%
\bibitem{Klaiman2008a}
S. Klaiman, U. G\"unther, and N. Moiseyev, Phys. Rev. Lett. \textbf{101}, 080402 (2008).
%
\bibitem{Klaiman2008b}
S. Klaiman and L. S. Cederbaum, Phys. Rev. A \textbf{78}, 062113 (2008).
%
\bibitem{Cartarius2007}
H. Cartarius, J. Main, and G. Wunner, Phys. Rev. Lett. \textbf{99}, 173003 (2007).
%
\bibitem{Cartarius2009}
H. Cartarius, J. Main, and G. Wunner, Phys. Rev. A \textbf{79}, 053408 (2009).
%
\bibitem{Lefebvre2009}
R. Lefebvre, O. Atabek, M. \v{S}indelka, and N. Moiseyev, Phys. Rev. Lett. \textbf{103}, 123003 (2009).
%
\bibitem{Gilary2013}
I. Gilary, A. A. Mailybaev, and N. Moiseyev, Phys. Rev. A \textbf{88}, 010102(R) (2013).
%
\bibitem{Dembowski2004}
C. Dembowski, B. Dietz, H.-D. Gr\"af, H. L. Harney, A. Heine, W. D. Heiss, and A. Richter,
Phys. Rev. E \textbf{69}, 056216 (2004).
%
\bibitem{Dietz2011}
B. Dietz, H. L. Harney, O. N. Kirillov, M. Miski-Oglu, A. Richter, and F. Sch\"afer, Phys. Rev. Lett. \textbf{106}, 150403 (2011).
%
\bibitem{Uzdin2010}
R. Uzdin and R. Lefebvre, J. Phys. B: At. Mol. Opt. Phys. \textbf{43}, 235004 (2010).
%
\bibitem{Kato1976}
T. Kato, \textit{Perturbation Theory for Linear Operators, 2nd ed.} (Springer, 1976).
%
\bibitem{Seyranian2003}
A. P. Seyranian and A. A. Mailybaev, \textit{Multiparameter Stability Theory with Mechanical Applications} (World Scientific, 2003).
%
\bibitem{Bhamathi1996}
G. Bhamathi and E. C. G. Sudarshan, Int. J. Mod. Phys. B \textbf{13 \& 14}, 1531 (1996).
%
\bibitem{Bohm1997}
A. Bohm, M. Loewe, S. Maxson, P. Patuleanu, C. P\"untmann, and M. Gadella, J. Math. Phys. \textbf{38}, 6072 (1997).
%
\bibitem{Antoniou1998}
I. E. Antoniou, M. Gadella, and G. P. Pronko, J. Math. Phys. \textbf{39}, 2459 (1998)
%
\bibitem{Hernandez2003}
E. Hern\'andez, A. J\'auregui, and A. Mondrag\'on, Phys. Rev. A \textbf{67}, 022721 (2003).
%
\bibitem{Berry2004}
M. V. Berry, Czech. J. Phys. \textbf{54}, 1039 (2004).
%
\bibitem{Heiss2004}
W. D. Heiss, Czech. J. Phys. \textbf{54}, 1091 (2004).
%
\bibitem{Heiss2012}
W. D. Heiss, J. Phys. A: Math. Theor. \textbf{45}, 444016 (2012).
%
\bibitem{Moiseyev1980}
N. Moiseyev and S. Friedland, Phys. Rev. A \textbf{22}, 618 (1980).
%
\bibitem{Gantmacher1959}
F. R. Gantmacher, \textit{Matrix Theory} (Chelsea, New York, 1959).
%
\bibitem{Horn2012}
R. A. Horn and C. R. Johnson, \textit{Matrix Analysis, 2nd ed.} (Cambridge University Press, 2012).
%
\bibitem{Nakanishi1972}
N. Nakanishi, Prog. Theor. Phys. Suppl. \textbf{51}, 1 (1972).
%
\bibitem{Gunther2007}
U. G\"unther, I. Rotter, and B. F. Samsonov, J. Phys. A: Math. Theor. \textbf{40}, 8815 (2007).
%
\bibitem{Demange2012}
G. Demange and E. M. Graefe, J. Phys. A: Math. Theor. \textbf{45}, 025303 (2012).
%
\bibitem{Rotter2009}
I. Rotter, J. Phys. A: Math. Theor. \textbf{42}, 153001 (2009).
%
\bibitem{Rotter2015}
I. Rotter and J. P. Bird, Rep. Prog. Phys. \textbf{78}, 114001 (2015)
%
\bibitem{Heiss2008}
W. D. Heiss, J. Phys. A: Math. Theor. \textbf{41}, 244010 (2008).
%
\bibitem{Nakanishi1958}
N. Nakanishi, Prog. Theor. Phys. \textbf{19}, 607 (1958).
%
\bibitem{Sudarshan1978}
E. C. G. Sudarshan, C. B. Chiu, V. Gorini, Phys. Rev. D \textbf{18}, 2914 (1978).
%
\bibitem{Petrosky1991}
T. Petrosky, I. Prigogine, and S. Tasaki, Physica A \textbf{173}, 175 (1991).
%
\bibitem{Bohm1993}
A. Bohm, \textit{Quantum Mechanics: Foundations and Applications, 3rd ed.} (Springer, 1993).
%
\bibitem{Hashimoto2015}
K. Hashimoto, K. Kanki, H. Hayakawa, and T. Petrosky, Prog. Theor. Exp. Phys. \textbf{2015}, 023A02 (2015).
%
\bibitem{Feshbach1958}
H. Feshbach, Ann. Phys. \textbf{5}, 357 (1958).
%
\bibitem{Feshbach1962}
H. Feshbach, Ann. Phys. \textbf{19}, 287 (1962).
%
\bibitem{Cohen-Tannoudji1992}
C. Cohen-Tannoudji, J. Dupont-Roc and G. Grynberg, \textit{Atom-Photon Interactions: Basic Processes and Applications} (Wiley, New York, 1992).
%
\bibitem{Petrosky1997}
T. Petrosky and I. Prigogine, Adv. Chem. Phys. \textbf{99}, 1 (1997).
%
\bibitem{Hatano2013}
N. Hatano, Fortschr. Phys. \textbf{61}, 238 (2013).
%
\bibitem{Hatano2014}
N. Hatano and G. Ordonez, J. Math. Phys. \textbf{55}, 122106 (2014).
%
\bibitem{Ordonez2016}
G. Ordonez and N. Hatano, arXiv:1610.01548.
%
\bibitem{Tolstikhin1998}
O. I. Tolstikhin, V. N. Ostrovsky, and H. Nakamura, Phys. Rev. A \textbf{58}, 2077 (1998).
%
\bibitem{Garmon2017}
S. Garmon and G. Ordonez, J. Math. Phys. \textbf{58}, 062101 (2017).
%
\bibitem{Bender2007}
C. M. Bender, Rep. Prog. Phys. \textbf{70}, 947 (2007).
%
\bibitem{Mostafazadeh2010}
A. Mostafazadeh, Int. J. Geom. Meth. Mod. Phys. \textbf{7}, 1191 (2010).
%
\bibitem{Bender2002}
C. M. Bender, M. V. Berry, and A. Mandilara, J. Phys. A: Math. Gen. \textbf{35}, L467 (2002).
%
\bibitem{Garmon2015}
S. Garmon, M. Gianfreda, and N. Hatano, Phys. Rev. A \textbf{92}, 022125 (2015).
%
\bibitem{Brody2016}
D. C. Brody, J. Phys. A: Math. Theor. \textbf{49}, 10LT03 (2016).
%
\bibitem{Tanaka2016}
S. Tanaka, S. Garmon, K. Kanki, and T. Petrosky, Phys. Rev. A \textbf{94}, 022105 (2016).
%
\bibitem{note}
For a prescription on the spectral decomposition of the effective Liouvillian,
see Ref.~\onlinecite{Petrosky1997}.
Formally the same method can be applied to the effective Hamiltonian,
but validity of the method should be investigated for each model individually.
%
\bibitem{chain}
In a general case of $M$-state coalescence ($M\ge 2$),
the Jordan chain is a relation among the eigenvector $|\phi_0\ket$ and $(M-1)$ pseudo-eigenvectors,
which we may denote by $|\phi_0^{(m)}\ket$ ($m=1,2,\cdots, M-1$).
%
\bibitem{Hinch1991}
E. J. Hinch, \textit{Perturbation Methods} (Cambridge University Press, 1991).
%
\bibitem{Garmon2012}
S. Garmon, I. Rotter, N. Hatano, and D. Segal, Int. J. Theor. Phys. \textbf{51}, 3536 (2012).
%
\bibitem{Brody2013}
D. C. Brody and E. M. Graefe, Entropy \textbf{15}, 3361 (2013).
%
\bibitem{Hashimoto2016a}
K. Hashimoto, K. Kanki, S. Tanaka, and T. Petrosky, Phys. Rev. E \textbf{93}, 022132 (2016).
%
\bibitem{Hashimoto2016b}
K. Hashimoto, K. Kanki, S. Garmon, S. Tanaka, and T. Petrosky, Prog. Theor. Exp. Phys. \textbf{2016}, 053A02 (2016).
%
\bibitem{Hashimoto2016c}
K. Hashimoto, K. Kanki, S. Tanaka, and T. Petrosky, 
in \textit{Non-Hermitian Hamiltonians in Quantum Physics}, F. Bagarello \textit{et al.} (eds.), Springer Proceedings in Physics 184, p. 263 (2016).
%
\bibitem{Kanki2016}
K. Kanki, K. Hashimoto, T. Petrosky, and S. Tanaka, 
in \textit{Non-Hermitian Hamiltonians in Quantum Physics}, F. Bagarello \textit{et al.} (eds.), Springer Proceedings in Physics 184, p. 289 (2016).
%
\bibitem{Kimura2002}
G. Kimura, K. Yuasa, and K. Imafuku, Phys. Rev. Lett. \textbf{89}, 140403 (2002).
%
\bibitem{Qian2000}
H. Qian and M. Qian, Phys. Rev. Lett. \textbf{84}, 2271 (2000).
%
\bibitem{Am-Shallem2015}
M. Am-Shallem, R. Kosloff, and N. Moiseyev, New J. Phys. \textbf{17}, 113036 (2015).
%
\bibitem{Am-Shallem2016}
M. Am-Shallem, R. Kosloff, and N. Moiseyev, Phys. Rev. A \textbf{93}, 032116 (2016).
%
\bibitem{5x5}
There is a $3\times 3$ matrix known as the Frobenius companion matrix\cite{Horn2012} which has the cubic function in Eq.~\eqref{cubic} as its characteristic polynomial as well as its minimal polynomial.
Similarly,
a $5\times 5$ matrix is associated with the quintic function in Eq.~\eqref{quintic}.
Each of the matrices is similar to a matrix with a Jordan block at the corresponding exceptional point.
The matrix of this kind lends itself to solving the non-linear eigenvalue problem of the effective Hamiltonian by numerical diagonalization,
although it has no direct physical meaning.
%
\end{thebibliography}
\end{document}